\documentclass{PoS}
\usepackage[utf8]{inputenc}
\usepackage{dsfont}
\usepackage{physics}
\usepackage{nicefrac}
\usepackage{mathtools}
\usepackage{tikz}

\newcommand*{\dpath}[1]{\!D#1\!\mathop{}}
\newcommand*{\seff}{S_{\mathrm{eff}}}
\renewcommand{\vec}[1]{\boldsymbol{\mathbf{#1}}}

\usetikzlibrary{positioning}
\tikzset{
  graph style/.style={
    baseline={([yshift=-.5ex]current bounding box.center)},
    node distance=.5,
  },
  skeleton node/.style={
    fill,circle,inner sep=0pt,minimum size=3.5pt
  },
  skeleton bond/.style={
    draw,
  },
}
\makeatletter
\tikzset{
  position/.style args={#1 degrees from #2}{
    at=(#2.#1), anchor=#1+180, shift=(#1:\tikz@node@distance)
  }
}
\makeatother

\title{Large $N_c$ behaviour of lattice QCD in the heavy dense regime}

\ShortTitle{Large $N_c$ behaviour of lattice QCD in the heavy dense regime}

\author{Owe Philipsen and \speaker{Jonas Scheunert}\\
  Institut für Theoretische Physik, Goethe-Universtität Frankfurt am Main\\
  Max-von-Laue-Str. 1, 60438 Frankfurt am Main, Germany
  E-mail: \email{scheunert,philipsen@th.physik.uni-frankfurt.de}}

\abstract{Combining strong coupling and hopping expansion one can derive a
  dimensionally reduced effective theory of lattice QCD. This theory has a
  reduced sign problem, is amenable to analytic evaluation and was successfully
  used to study the cold and dense regime of QCD for sufficiently heavy quarks.
  We show results from the evaluation of the effective theory for arbitrary
  $N_c$ up to $\kappa^4$. The inclusion of gauge corrections is also
  investigated. We find that the onset transition to finite baryon number
  density steepens with growing $N_c$ even for $T \neq 0$. This suggests that
  in the large $N_c$ limit the onset transition is first order up to the
  deconfinement transition. Beyond the onset, the pressure is shown to scale
  as $p \sim N_c$ through three orders in the hopping expansion, which is
  characteristic for a phase termed quarkyonic matter in the literature.}

\FullConference{37th International Symposium on Lattice Field Theory - Lattice2019\\
  16-22 June 2019\\
  Wuhan, China}

\begin{document}

\section{Introduction}

Large parts of the phase structure of QCD for finite baryon chemical potential
are still unknown. This is due to a sign problem which prohibits simulations
using importance sampling at finite density. Various approximate methods that
extend the Monte-Carlo method to finite densities exist, they are, however,
limited to low densities or high temperature $\mu_B/T \lessapprox 3$. In this
region, no sign of criticality or a first order phase transition has been found
using these methods \cite{Ratti:2019tvj}. Regarding non-perturbative continuum
approaches, based on the functional renormalisation group a critical endpoint
has recently been reported at $(T_\mathrm{CEP},\mu_{B,\mathrm{CEP}}) =
(107,635)\;\mathrm{MeV}$ \cite{Fu:2019hdw}, however with systematic
uncertainties growing large in the density region around the critical endpoint.

It is therefore interesting to study effective lattice theories with a sign
problem that is mild enough to also allow simulations in the cold and dense
region, or a completely analytic evaluation. One approach to obtain such a
theory is to integrate the spatial gauge links via a combined strong coupling
and hopping expansion, leading to a 3-dimensional effective theory of lattice
QCD. This theory is valid for sufficiently heavy quarks on reasonably fine
lattices and can be used to investigate the cold and dense regime around the
onset to baryon matter. It can be simulated using reweighting and complex
Langevin \cite{Fromm:2011qi,Langelage:2014vpa} and can be evaluated
analytically using a linked cluster expansion known from statistical mechanics
\cite{Glesaaen:2015vtp}.
Here we extend this analytic evaluation to arbitrary $N_c$ up to order
$\kappa^4$. This enables us to make a connection to another approach to dense
and cold QCD, namely that based on large $N_c$ considerations. Specifically, in
\cite{McLerran:2007qj} the authors argue for the existence of a phase termed
quarkyonic matter, which has both quark and baryon-like properties and is
characterised by pressure scaling as $p_c \sim N_c$.

\section{The effective theory for general $N_c$}

The effective theory is based on lattice QCD with standard Wilson fermions.
Finite temperature is implemented by a compact euclidean time dimension with
$N_{\tau}=\nicefrac{1}{a T}$ slices and (anti-) periodic boundary conditions for
(fermions) bosons. Finite chemical potential is introduced by a factor of
$\exp((-) a \mu)$ in front of temporal (anti-) quark hops. Integrating out the
Grassmann fields and spatial gauge links $U_i \in SU(N_c)$ leads to an
effective theory which only depends on temporal links $U_0 \in SU(N_c)$:
\begin{align}
  Z 
  & =
  \int \dpath{U} \dpath{\Psi} \dpath{\bar{\Psi}} 
  e^{-S_G[U]-S_f^{(W)}[U,\Psi,\bar{\Psi}]} \\
  & \eqqcolon
  \int \dpath{U_0} e^{-\seff[U_0]}
  =
  \int \dpath{W} e^{-\seff[W]} \\
  \Rightarrow \seff[U_0]
  & =
  \label{eq:efft-definition}
  -\log(\int\dpath{U_i}\dpath{\Psi}\dpath{\bar{\Psi}}
  e^{-S_G[U]-S_f^{(W)}[U,\Psi,\bar{\Psi}]})
\end{align}
Gauge invariance necessitates that the dependence of the effective action on
temporal links is in terms of traces of powers of Wilson lines $W$,
\begin{equation}
  W(\vec{n}) = \prod_{\tau=1}^{N_\tau} U_0(\tau,\vec{n}).
\end{equation}
At this point the effective theory is uniquely determined by
eq. \eqref{eq:efft-definition}. However, the integration of the spatial links
lead to interactions to all distances, so in practice truncations have to be
introduced.
Here, we use a character (resummed strong coupling) expansion for the gauge
part and a hopping expansion for the fermion part of the action. Truncating
these expansions at a certain order enables an analytic evaluation of the path
integral to that order. We start in the strong coupling limit, setting the
lattice gauge coupling $\beta=\frac{2 N_c}{g^2} = 0$, and discuss the
systematics of the hopping expansion. This is done in an alternative approach
in comparison to earlier publications. It is based on integrating the spatial
gauge fields before integrating Grassmann valued fields, well known from the
approach to staggered fermions in the strong coupling limit
\cite{Rossi:1984cv,deForcrand:2014tha}. This makes the systematic inclusion of
low $N_\tau$-contributions, which were neglected in \cite{Glesaaen:2015vtp} and
are relevant to get the correct $N_c$ scaling, easier.

\subsection{Systematics of the hopping expansion}

At $\beta=0$, the link integration factorizes and splitting temporal and
spatial hops results in 
\begin{equation}
  \label{eq:exp-S_eff-split}
    -\seff[U_0] = 
    \log\biggl[
    \int\dpath{\Psi}\dpath{\bar{\Psi}}e^{\bar{\Psi}(-1+T[U_0])\Psi}
    \prod_{n\in\Lambda}\prod_{i=1}^3
    \int\dd{U_i(n)}
  e^{\kappa\tr(J_i(n)U_i(n) + U_i^\dagger(n)K_i(n))}\biggr].
\end{equation}
Here, $\Lambda$ refers to the $N_\tau \times N_s$ lattice, $\kappa$ is the
hopping expansion parameter and is related to the bare quark mass via
$\kappa=\nicefrac{1}{(2am_q+8)}$, $T$ is the temporal hopping matrix
\begin{equation}
  T[U_0](n,m) = 
  \kappa e^\mu (1-\gamma_0) U_0(n)
  \delta_{n_0,n_0+1} \delta_{\vec{n},\vec{m}} +
  \kappa e^{-\mu} (1+\gamma_0) U_0(n)^\dagger
  \delta_{n_0,n_0+1} \delta_{\vec{n},\vec{m}},
\end{equation}
and the spatial hops are encoded in $J$ and $K$ in the following way:
\begin{align}
  J_i(n)_{a b}
  & = 
  \bar{\Psi}(n)_{\alpha,b}^{f}
  (1 - \gamma_i)_{\alpha\beta}
  \Psi(n+\vec{e}_i)_{\beta,a}^{f} \\
  K_i(n)_{a b}
  & = 
  \bar{\Psi}(n+\vec{e}_i)_{\alpha,b}^{f}
  (1 + \gamma_i)_{\alpha\beta}
  \Psi(n+\vec{e}_i)_{\beta,a}^{f}.
\end{align}
In $\bar{\Psi}^f_{\alpha,b}$, $f$ refers to flavour, $\alpha$ to Dirac and $b$
to $SU(N_c)$ indices. All flavours are assumed to be degenerate. The
single site integral in eq. \eqref{eq:exp-S_eff-split} was given for $U(N_c)$ in
\cite{Bars:1980yy} in terms of irreducible characters $\chi_r$ of the general linear
group. Although we are interested in $SU(N_c)$ here, the only difference to
$U(N_c)$ is due to spatial baryon hoppings, which have a prefactor
$\kappa^{k N_c}$, meaning they are suppressed for large $N_c$. Since we are
ultimately interested in large $N_c$ considerations, we neglect them here,
which means we can use the $U(N_c)$ results for the spatial integration.
As a result one obtains a multinomial $M$ of the $\Psi$s with which the
effective action can be represented as
\begin{align}
  \label{eq:exp-S_eff-U_i-integrated}
  \begin{split}
    -\seff[U_0]
    & =
    \log\biggl[\int\dpath{\Psi}\dpath{\bar{\Psi}}e^{\bar{\Psi}(-1+T[U_0])\Psi} \\
    & \qquad \times
    \prod_{n\in\Lambda}\prod_{i=1}^3
    \qty(1 + M(\Psi(n),\Psi(n+\vec{e}_i),\bar{\Psi}(n),\bar{\Psi}(n+\vec{e}_i)))
  \biggr].
  \end{split}
\end{align}
After expanding the product, the integration can be done using Wick's theorem.
The corresponding propagator $(1-T[U_0])^{-1}$ has been obtained in
\cite{Langelage:2014vpa} and
is diagonal with respect to its spatial arguments but not with respect to
its temporal arguments. Therefore, to organize the expansion we write the
lattice as a product of time slices of spatial lattices
$\Lambda=\Lambda_\tau\times\Lambda_s$ and rewrite the product in
eq. \eqref{eq:exp-S_eff-U_i-integrated} to
\begin{align}
  \begin{split}
    \prod_{n\in\Lambda}\prod_{i=1}^3
    &
    \qty(1 + M(\Psi(n),\Psi(n+\vec{e}_i),\bar{\Psi}(n),\bar{\Psi}(n+\vec{e}_i))) \\
    & =
    \prod_{\vec{n}\in\Lambda_s}\prod_{i=1}^3
    \prod_{\tau\in\Lambda_\tau}
    \qty(1 + 
    M(\Psi(\tau,\vec{n}),
    \Psi(\tau,\vec{n}+\vec{e}_i),
    \bar{\Psi}(\tau,\vec{n}),
    \bar{\Psi}(\tau,\vec{n}+\vec{e}_i)))
  \end{split} \\
  \label{eq:graph-product}
  & \eqqcolon
  \prod_{\vec{n}\in\Lambda_s}\prod_{i=1}^3
  \qty(
  1 +
  \begin{tikzpicture}[graph style]
    \node[skeleton node] (n1) {};
    \node at (n1) [left = -1mm of n1] {$\vec{n}$};
    \node[skeleton node] (n2) [right=of n1] {};
    \node at (n2) [right = -1mm of n2] {$(\vec{n}+\vec{e}_i)$};
    \draw[skeleton bond] (n1) -- (n2);
  \end{tikzpicture}),
\end{align}
The expansion of the product can then be represented
as a sum of connected and disconnected subgraphs of the spatial lattice, with the
graph in eq. \eqref{eq:graph-product} as an elementary building block of more
complicated graphs. Denoting by $\Phi$ the evaluation of a graph according to
\begin{equation}
  \Phi\qty(
  \begin{tikzpicture}[graph style]
    \node[skeleton node] (n1) {};
    \node at (n1) [left = -1mm of n1] {$\vec{n}$};
    \node[skeleton node] (n2) [right=of n1] {};
    \node at (n2) [right = -1mm of n2] {$(\vec{n}+\vec{e}_i)$};
    \draw[skeleton bond] (n1) -- (n2);
  \end{tikzpicture}
  ) = 
  \frac{
    \int\dpath{\Psi}\dpath{\bar{\Psi}}e^{\bar{\Psi}(-1+T[U_0])\Psi}
    \begin{tikzpicture}[graph style]
      \node[skeleton node] (n1) {};
      \node at (n1) [left = -1mm of n1] {$\vec{n}$};
      \node[skeleton node] (n2) [right=of n1] {};
      \node at (n2) [right = -1mm of n2] {$(\vec{n}+\vec{e}_i)$};
      \draw[skeleton bond] (n1) -- (n2);
  \end{tikzpicture}}
  {
  \int\dpath{\Psi}\dpath{\bar{\Psi}}e^{\bar{\Psi}(-1+T[U_0])\Psi}},
\end{equation}
the evaluation factorizes over disconnected graphs.
Then, the moment-cumulant formalism \cite{Ruelle:1969,Montvay:1994cy} can be
used to expand the logarithm in eq. \eqref{eq:exp-S_eff-U_i-integrated}, resulting in
an expansion in connected clusters of graphs
\begin{equation}
  \label{eq:efft-strong-coupling-evaluated}
  -\seff[U_0] =
  \sum_{\vec{n}\in\Lambda_s} \log(z_0(W(\vec{n}))) +
  \sum_{n=1}^\infty \sum_{g_1,\ldots,g_n\in\mathcal{G}_c}
  \frac{1}{n!} \qty[g_1,\ldots,g_n] \Phi(g_1) \dotsm \Phi(g_n).
\end{equation}
In this equation, $\qty[\ldots]$ takes integer values that are non-vanishing exactly
when all the graphs in its argument form a cluster of connected graphs.
While eq. \eqref{eq:efft-strong-coupling-evaluated} is exact, in practice one truncates
the infinite sum by including only contributions up to a certain power in $\kappa$.

\subsection{Evaluation of the effective theory}
Evaluating the effective theory eq. \eqref{eq:efft-strong-coupling-evaluated}
along the lines of \cite{Glesaaen:2015vtp}, the free energy to
$\mathcal{O}(\kappa^2)$ reads
\begin{equation}
  -f =
  \log(z_0(h_1)) -
  6 N_f \frac{\kappa^2 N_\tau}{N_c} \qty(\frac{z_{11}(h_1)}{z_0(h1)}),
\end{equation}
with the $SU(N_c)$ integrals
\begin{align}
  z_0 & = \int\limits_{SU(N_c)} \dd{W} \det(1 + h_1 W)^{2 N_f}, \\
  z_{11} & = \int\limits_{SU(N_c)} \dd{W} \det(1 + h_1 W)^{2 N_f}
  \tr\qty(\frac{h_1 W}{1 + h_1 W}),
\end{align}
where $h_1 = (2 \kappa e^{a\mu})^{N_\tau} = e^{\frac{\mu-m}{T}}$ and $a
m=-\log(2\kappa)$ is the leading order expression 
of the constituent quark mass of a baryon in lattice units. We have also neglected
all contributions containing factors of $\bar{h}_1=(2 \kappa
e^{-a\mu})^{N_\tau}$, which is justified because we first want to consider low
temperatures. The integrals can be solved in the Polyakov gauge
\cite{Nishida:2003fb}, for details and the $\mathcal{O}(\kappa^4)$ contribution
to the free energy we refer to \cite{Philipsen:2019qqm}.

\section{Large $N_c$ behaviour of the effective theory}
Having obtained the free energy in the previous section, we can investigate
thermodynamic observables in the $\mu_B=3\mu$ and $T$-plane for general and
large $N_c$. To illustrate the general strategy, consider for $N_f=1$ the
$\kappa^2$ correction to the pressure, which reads
\begin{equation}
  a^4 p_1 =
  -6 \kappa^2 
  \frac{(\frac{1}{2}N_c(N_c+1)h_1^{N_c} + N_c h_1^{2 N_c})^2}
  {N_c(1+h_1^{N_c}(1+N_c)+h_1^{2N_c})^2}.
\end{equation}
For $h_1<1$, the $h_1^{N_c}$ factors are strongly suppressed for
$N_c\rightarrow\infty$ and a Taylor expansion around $h_1^{N_c} = 0$ results in
\begin{equation}
  a^4 p_1 =
  -\frac{3}{2} \kappa^2 N_c (N_c+1)^2 h_1^{2N_c} + \mathcal{O}(h_1^{3 N_c}) \sim
  -\frac{3}{2} \kappa^2 N_c^3 h_1^{2 N_c}.
\end{equation}
Similarly, for $h_1>1$, expanding about $1/h_1^{N_c}=0$ gives
\begin{align}
  a^4 p_1 =
  -6 \kappa^2 N_c + \mathcal{O}(1/h_1^{N_c}) \sim
  -6 \kappa^2 N_c.
\end{align}
Using this strategy one obtains for $N_f=2$ degenerate flavours for the
pressure and baryon density:
\begin{align}
  p 
  & \sim 
  \begin{cases}
    \frac{1}{6 a^4 N_{\tau}} N_c^3 h_1^{N_c} -
    \kappa^2 \frac{1}{48 a^4} N_c^7 h_1^{2 N_c} + 
    \kappa^4 \frac{3 N_\tau \kappa^4}{800 a^4} N_c^8 h_1^{2 N_c} +
    \mathcal{O}(\kappa^6),
    & \text{ if } h_1<1, \\
    \frac{4 \log(h_1)}{N_{\tau} a^4} N_c - 
    \kappa^2 \frac{12}{a^4} N_c  + 
    \kappa^4 \frac{198}{a^4} N_c +
    \mathcal{O}(\kappa^6),
    & \text{ if } h_1>1,
  \end{cases} \\
  n_B
  & \sim
  \begin{cases}
    \frac{1}{6 a^3} N_c^3 h_1^{N_c} - 
    \kappa^2 \frac{N_{\tau}}{24 a^3} N_c^7 h_1^{2 N_c} +
    \kappa^4 \frac{(9 N_\tau+1)N_\tau}{1200 a^3} N_c^8 h_1^{2 N_c} +
    \mathcal{O}(\kappa^6),
    & \text{ if } h_1 < 1, \\
    \frac{4}{a^3} - 
    \kappa^2 \frac{N_{\tau}}{a^3} \frac{N_c^4}{h_1^{N_c}} -
    \kappa^4 \frac{(59 N_\tau - 19) N_\tau}{20 a^3} \frac{N_c^5}{h_1^{N_c}} +
    \mathcal{O}(\kappa^6),
    & \text{ if } h_1 >1. \\
  \end{cases}
\end{align}
One can make several observations based on these formulas. For $h_1 <
1$, i.\,e.\ before the baryon onset, the observables are exponentially suppressed for
$N_c \rightarrow \infty$, for all $N_\tau$.

For the baryon density one then has a first order jump to the lattice
saturation density $a^3 n_B^{\mathrm{sat}} = 2 N_f$ starting at $h_1>1$. The
saturation density is a discretization artefact determined by the leading order
contribution in the hopping expansion, i.\,e.\ the static determinant. Higher
orders in $\kappa$ do not contribute. The boundary $h_1 =1$ does not depend on
$T$, therefore this transition should always be first order, until one reaches
another discontinuity. Note however, that we neglected $\bar{h}_1$, so for
higher temperatures this might be questionable. The inclusion of $\bar{h}_1$
was discussed in \cite{Philipsen:2019qqm}, leaving the qualitative observations
unchanged.

For the pressure we observe that it scales linearly in $N_c$ after
the onset, a property that characterises quarkyonic matter
\cite{McLerran:2007qj}. Just like for the baryon density, the leading
order is determined by lattice saturation, which is a discretization artefact.
However, for the pressure the linear $N_c$ scaling is also observed in all
computed higher order terms, which do not contribute to saturation. Therefore,
this suggests to hold to all orders in the hopping expansion, and
then would be a genuine effect beyond saturation for all current
quark masses.

\subsection{Gauge corrections and 't Hooft scaling}
So far, all quantities were analysed in the strong coupling limit $\beta=0$.
The continuum studies we are interested in employ the 't Hooft limit
\cite{tHooft:1973alw}, which is defined by taking $N_c \rightarrow \infty$ with
$\lambda_H \coloneqq g^2 N_c$ kept fixed. Therefore, in order to take the same
limit, we have to include corrections due to the gauge part of the action,
which are of course anyway important in order to make connections to continuum
results.
Many aspects related to the gauge sector of the effective theory for general
$N_c$ were discussed in \cite{Christensen:2013xea}. The strong coupling
expansion around $\beta=0$ is done by a character expansion. Here we only include
leading gauge corrections from the character expansion to the fermion
determinant. This works the same as in \cite{Langelage:2014vpa}, the only
difference being that the expansion coefficient of the fundamental character
$u(\beta)$ has to be replaced by its proper generalisation, which can be found,
for example, in \cite{Drouffe:1983fv}.
To $\mathcal{O}(\kappa^2)$ the free energy then reads
\begin{equation}
  -f = \ln(z_0(h_1)) + 
    \frac{\kappa^2 N_\tau}{N_c}\left[1+2 \frac{u-u^{N_\tau}}{1-u}\right](-6 N_f)
    \frac{z_{11}(h_1)}{z_0(h_1)}^2,
\end{equation}
with a modified $h_1$ due to gauge corrections
\begin{equation}
  h_1 = 
  (2 \kappa)^{N_\tau} e^{a \mu N_\tau}
  \exp\left[6 N_\tau \kappa^2 \frac{u-u^{N_\tau}}{1-u}\right].
\end{equation}
In taking the 't Hooft limit by keeping $\lambda_H=\frac{2N_c^2}{\beta}$ fixed as
$N_c\rightarrow\infty$, one can use that for $\lambda_H>1$ \cite{Gross:1980he}
\begin{equation}
  u(\beta) = \frac{1}{\lambda_H}
\end{equation}
in this limit. Therefore, these corrections shift the point on the chemical
potential axis where $h_1=1$ and besides that only modify the asymptotic
behaviour of the observables by a constant $\sim N_c^0$. In
\cite{Philipsen:2019qqm}, we also discussed the inclusion of the leading order
contribution of the pure gauge sector to the effective theory and similarly
observed no qualitative changes in the large $N_c$ behaviour.
One should, however, mention one important caveat. Based on an analysis of QCD
in $1+1$ dimensions, the interchange of the $N_c\rightarrow$ and the strong
coupling expansion was declared to be ``highly suspicious'' in
\cite{Gross:1980he}, which we did so far. Furthermore, the fact that
the density at the onset transition immediately jumps to lattice saturation
indeed suggests that to get results for continuum physics one should take the
continuum limit before taking the large $N_c$ limit. This has been investigated
further in \cite{Philipsen:2019qqm}, where the linear $N_c$ scaling of the
pressure was observed before saturation and this behaviour was at least
observed to be stable when the lattice is made finer.

\section{Conclusion}
We have investigated the large $N_c$ behaviour of an effective theory of
lattice QCD for heavy quarks, which is based on a combined strong coupling and
hopping expansion. We have found that in the baryon condensed phase the
pressure scales linearly in $N_c$ and the onset transition becomes first-order.
These results are in agreement with the proposed large $N_c$ phase diagram in
\cite{McLerran:2007qj}.

\acknowledgments

The authors acknowledge support by the Deutsche Forschungsgemeinschaft (DFG)
through the grant CRC-TR 211 ``Strong-interaction matter under extreme
conditions'' and by the Helmholtz International Center for FAIR within the LOEWE
program of the State of Hesse.

\bibliographystyle{JHEP}
\bibliography{references}

\end{document}